%% file: bare_jrnl.tex
\documentclass[journal]{IEEEtran}
\usepackage{amsmath,amssymb}
\usepackage{mwe}
\usepackage{hyperref}
\usepackage[switch]{lineno}
\usepackage{multicol}
\usepackage{textcomp}
\usepackage{tikz}
\usepackage{upgreek}
\usepackage{siunitx}
\usepackage{adjustbox}

\usepackage{cite}

\ifCLASSINFOpdf
\else
\fi
\ifCLASSOPTIONcompsoc
 \usepackage[caption=false,font=normalsize,labelfont=sf,textfont=sf]{subfig}
\else
 \usepackage[caption=false,font=footnotesize]{subfig}
\fi
\newcommand{\red}[1]{\textcolor{black}{#1}}

\begin{document}
%
\title{Large-Scale Detector Testing for the GAPS Si(Li) Tracker}
%
%
%

\author{Mengjiao~Xiao,
        Achim~Stoessl,
        Brandon~Roach,
        Cory~Gerrity,
        Ian~Bouche,
        Gabriel~Bridges,
        Philip~von~Doetinchem,
        Charles~J.~Hailey,
        Derik~Kraych,
        Anika~Katt,
        Michael~Law,
        Alexander~Lowell,
        Evan~Martinez,
        Kerstin~Perez,
        Maggie~Reed,
        Chelsea~Rodriguez,
        Nathan~Saffold,
        Ceaser~Stringfield,
        Hershel~Weiner,
        Kelsey~Yee


\thanks{M.~Xiao, B.~Roach, I.~Bouche, A.~Katt, K.~Perez, and K.~Yee were with the Massachusetts Institute of Technology, Cambridge, MA 02139, USA.}
\thanks{A.~Stoessl, C.~Gerrity, P.~von~Doetinchem, and H.~Weiner were with the University of Hawai'i at M{\=a}noa, Honolulu, HI 96822, USA.} 
\thanks{G.~Bridges, C.~Hailey, D.~Kraych,  M.~Law, E.~Martinez, K.~Perez, M.~Reed, C.~Rodriguez, N.~Saffold, and C.~Stringfield were with Columbia University, New York, NY 10027, USA.} 
\thanks{A. Lowell was with the Space Sciences Laboratory, University of California, Berkeley, Berkeley, CA 94720, USA.}
\thanks{\red{GAPS is supported in the U.S. by the NASA APRA program (Grant Nos. NNX17AB44G,
NNX17AB46G, and NNX17AB47G), in Japan by the
JAXA/ISAS Small Science Program FY2017, and in Italy
by the ASI/INFN
agreement No. 2018-22-HH.0: ``Partecipazione italiana al
GAPS --- General AntiParticle Spectrometer.'' M. Xiao, B. Roach, and K. Perez are supported by Heising-Simons Foundation award 2018-0766. C. Gerrity is supported by NASA under
award No. 80NSSC19K1425 of the Future Investigators in NASA Earth and Space Science and Technology (FINESST) program. K. Yee is supported through the U.S. National Science Foundation Graduate Research Fellowship under grant 2141064.}}
}

\maketitle


\begin{abstract}
Lithium-drifted silicon [Si(Li)] has been used for decades as an ionizing radiation detector in nuclear, particle, and astrophysics experiments, though such detectors have frequently been limited to small sizes ($\boldmath{\text{few\,cm}^2}$) and cryogenic operating temperatures. The 10-cm-diameter Si(Li) detectors developed for the General Antiparticle Spectrometer (GAPS) balloon-borne dark matter experiment are novel particularly for their requirements of low cost, large sensitive area ($\boldsymbol{{\sim} 10\text{\,m}^2}$ for the full 1440-detector array), high temperatures (near --40$\boldsymbol{\,^\circ}$C), and energy resolution below 4 keV FWHM for 20--100-keV x-rays. Previous works have discussed the manufacturing, passivation, and small-scale testing of prototype GAPS Si(Li) detectors. Here we show for the first time the results from detailed characterization of over 1100 flight detectors, illustrating the consistent intrinsic low-noise performance of a large sample of GAPS detectors. This work demonstrates the feasibility of large-area and low-cost Si(Li) detector arrays for next-generation astrophysics and nuclear physics applications.    
\end{abstract}



%
\IEEEpeerreviewmaketitle

\section{Introduction}
%
%
%
%
\IEEEPARstart{F}{irst} employed as an ionizing radiation detector in the 1950s, lithium-drifted silicon [Si(Li)] detectors have since found wide use in a variety of nuclear, particle, and astrophysics applications as charged-particle and photon detectors (e.g.,~\cite{Ahmad:1974,Krimigis:1977a,Giacconi:1979,Gerbier:1990,Roberts:1995,Stone:1998a,Popeko:2008,Gurov:2010,PhysRevLett.104.123001,Aseev:2011,Derbin:2012}). The introduction of lithium ions into p-type silicon compensates impurities in the crystal lattice, producing nearly-intrinsic regions up to a few millimeters in thickness. Early Si(Li) detectors were generally limited to active areas of a few cm$^2$ by the size and purity of available monocrystalline silicon wafers, by the increase in capacitance with detector area, and by the uniformity in lithium drifting across the detector (e.g., ~\cite{Yoshimori:1975a}). In time, improvements in production methods enabled the development of larger Si(Li) detectors with energy resolutions suitable for x-ray spectroscopy (e.g.,~\cite{PEHL1986519,FONG1982623}).
\par Si(Li) detector technology is ideal to meet the unique performance requirements of the General Antiparticle Spectrometer (GAPS) experiment. GAPS is a balloon-borne particle tracker and x-ray spectrometer, optimized to detect low-energy (${<}0.25$ GeV/nucleon) cosmic antinuclei (i.e., antiprotons, antideuterons, and antihelium) resulting from dark-matter decay and/or annihilation in the Milky Way~\cite{Mori:2001dv,Hailey:2009a,Aramaki:2014oda,Hailey:2013a,Aramaki:2015laa,vonDoetinchem:2020vbj,Saffold:2020axg,GAPS:2022ncd}. At the heart of the GAPS payload and particle identification scheme is a novel exotic-atom technique. An incoming antinucleus first passes through plastic-scintillator time-of-flight (TOF) panels, providing velocity and energy-deposition (d$E$/d$x$) information. The antinucleus then experiences d$E$/d$x$ losses in the Si(Li) tracker before coming to rest and being captured by an atomic nucleus in the Si(Li) detectors or surrounding support materials. This exotic atom subsequently de-excites, emitting a cascade of characteristic x-rays whose energies are related to the reduced mass and nuclear charge of the nucleus-antinucleus pair. Finally, the nucleus and antinucleus annihilate, producing a shower of pions and protons whose multiplicity scales with the mass of the antinucleus. This combination of signatures provides strong rejection power against the abundant protons, alpha particles, and other non-antimatter cosmic-ray particles.
\par Deploying this particle-identification technique on a balloon experiment such as GAPS presents several unique challenges. First, a large active detector area (${\sim}$10 m$^2$) is required to search for rare signals such as antideuterons or antihelium. Second, the detectors must be sufficiently thick to serve as a stopping target for low-energy antinuclei, while also being sufficiently thin for characteristic x-rays with energies ${\sim}$20--100 keV to escape. Third, mass and power constraints prevent the use of a cryostat large enough to contain the full tracker, so the GAPS detectors must be operable near --40$\,^\circ\text{C}$, achieved with a novel oscillating heat pipe thermal system~\cite{HideyukiFUKE2016,OKAZAKI201820}. These requirements motivate the use of 10-cm-diameter and 2.5-mm-thick Si(Li) detectors for the GAPS tracker, with each detector segmented into eight strips. Previous works have described the fabrication, passivation, and testing of several prototype detectors~\cite{Perez:2018akh,Kozai:2019xlp,Rogers:2019avj,Kozai:2021apo,Saffold:2021uqx}. 
\par In this paper, we describe the low-noise and low-temperature characterization of the more than 1100 Si(Li) detectors that will be used for the first GAPS flight. Before integrating the Si(Li) detectors into GAPS, we must characterize the performance of each detector at the operating temperatures expected in flight. These tests serve several purposes. First, we must verify that the detector is operable (i.e., that its current draw will not saturate the flight high-voltage power system). Second, we must characterize the x-ray energy resolution of each strip. From these x-ray measurements we construct noise models (i.e., energy resolution as a function of peaking time) for each detector, allowing us to verify that the detector is performing as expected and to predict the detector performance using the flight readout system. Third, we will use these measurements to determine the placement of detectors within the tracker, with higher-quality detectors being placed in the upper layers (where more anti-nuclei events are expected).  
\par In Sec.~\ref{sec:testing_system} we describe the custom low-noise, low-temperature testing system that was developed to allow fast characterization of this large sample of detectors. In Sec.~\ref{sec:testing_flow}, we describe our procedures for detector processing, testing, and data acquisition. In Sec.~\ref{sec:analysis_results}, we characterize this large sample of detectors using a physically-motivated noise model. We conclude in Sec.~\ref{sec:conclusion}, demonstrating that large-area and low-noise Si(Li) detectors can be economically mass-produced and tested. 


 




%
%

\begin{figure}[h]
\centering

\subfloat{\includegraphics[width=0.76\columnwidth]{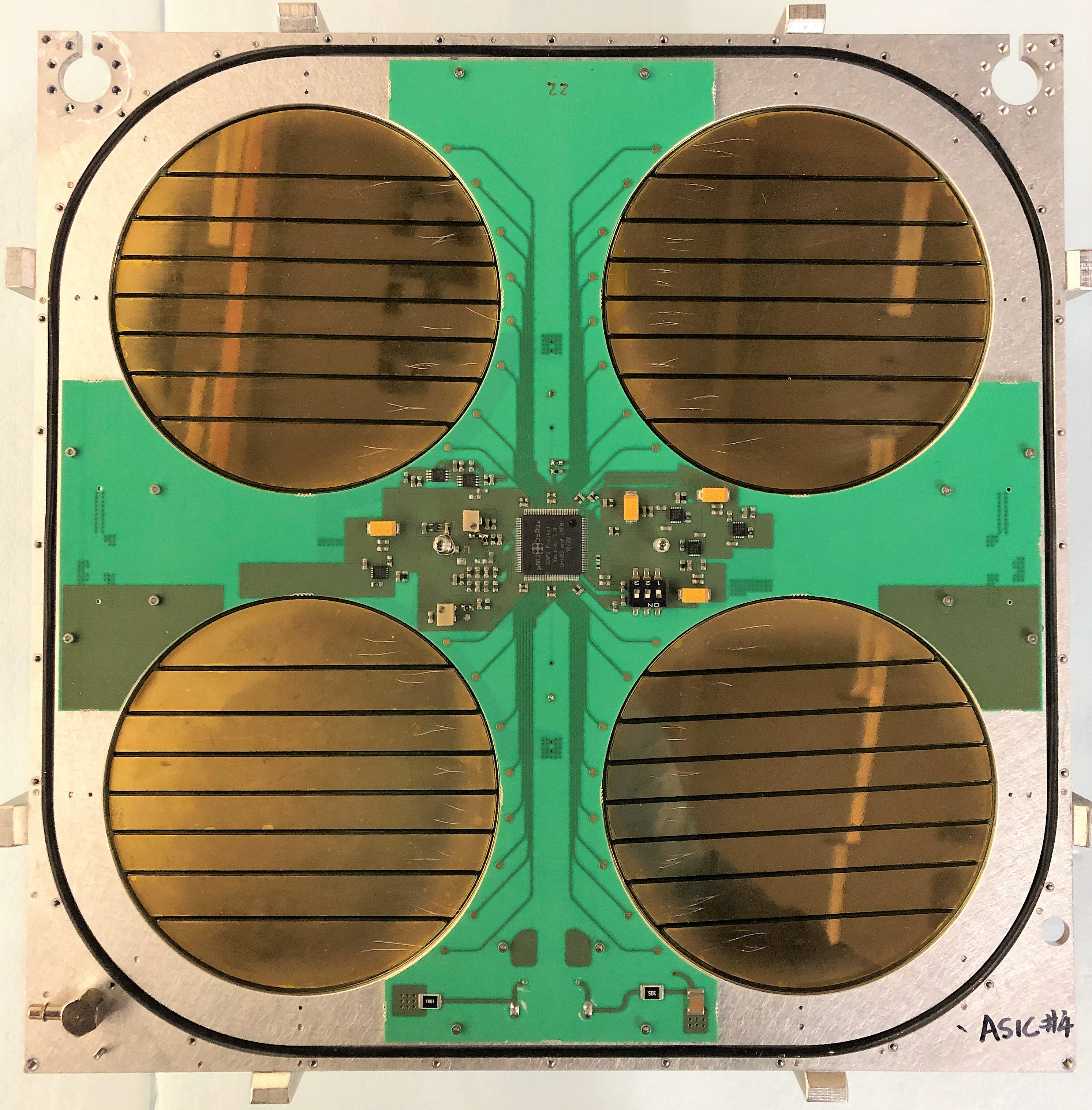}}

\vspace{0.3cm}

\subfloat{\includegraphics[width=0.75\columnwidth]{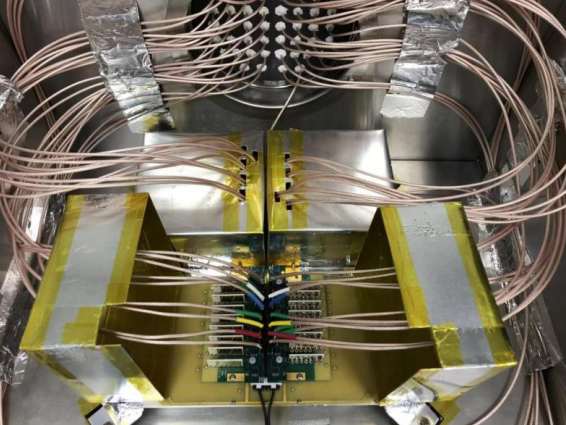}}

\vspace{0.3mm}

\subfloat{

\includegraphics[width=0.75\columnwidth]{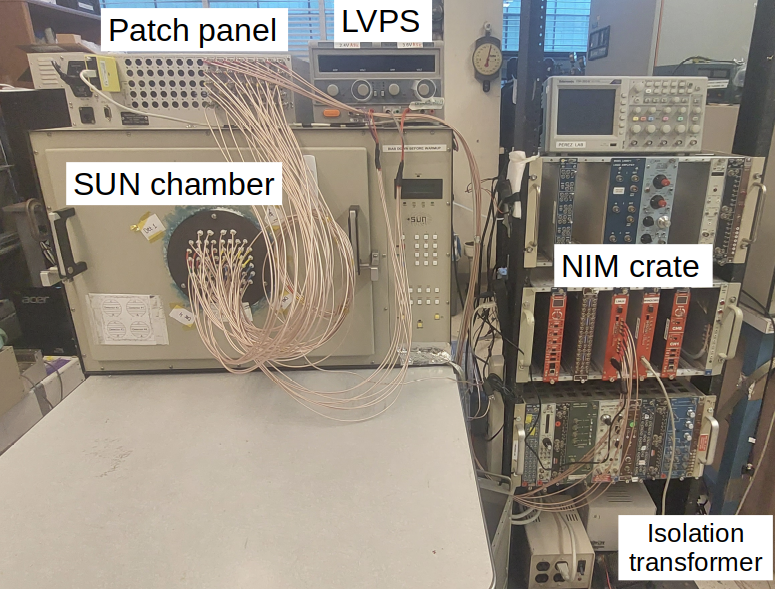}

}

\caption{\textbf{Top}: A GAPS flight module without the top polypropylene window. The FEB is the green cross-shaped board, in the center of which is the custom ASIC chip (black). \textbf{Middle}: A module mounted to the SUN chamber door for testing. Two Faraday cages have been pulled to the side to show the discrete preamplifier boards (green) mounted to the FR4 testing board (yellow). \textbf{Bottom}: The MIT Si(Li) testing system, with major components labeled. A system using identical components was constructed at UHM.}
\label{figure:module}
\end{figure}

\section{\label{sec:testing_system}{Detector Testing System}}

\subsection{GAPS Detector Module}
The basic functional unit of the GAPS Si(Li) tracker is the \textit{detector module} (hereafter simply \textit{module}) shown in Fig.~
\ref{figure:module}. Each module contains four Si(Li) detectors mounted into an aluminum frame, to which the detectors are grounded via their guard rings. Each detector is rigidly mounted into place using several fluorosilicone O-rings and G10 clamps. These clamps also secure a gold tab to the bottom ohmic contact of each detector to supply the bias voltage. Finally, we secure polypropylene windows on the top and bottom of each detector module, providing a sealed volume for the detectors. These windows (aluminized on their exterior surface to prevent metal flakes within the detector volume) allow us to shield the detectors from stray light, environmental contamination, and electromagnetic interference, and to flush the modules with dry N$_2$ gas during payload assembly and testing to suppress humidity. Each module also carries a \textit{front-end board} (FEB) with several major roles. First, each FEB carries a custom ASIC chip to which the Si(Li) strips are wirebonded, and which is responsible for processing and digitizing Si(Li) energy deposits over a large dynamic range of tens of keV to ${\gtrsim}$100 MeV~\cite{7286866,Scotti:2019lea,Manghisoni:2021,RE2023167617}. Second, the FEBs also provide the 250-V reverse-bias voltage to each detector. Lastly, the FEBs facilitate a power and communication link between the flight computer and the ASICs.
\par
\begin{figure}[t]
    \includegraphics[width=1\columnwidth]{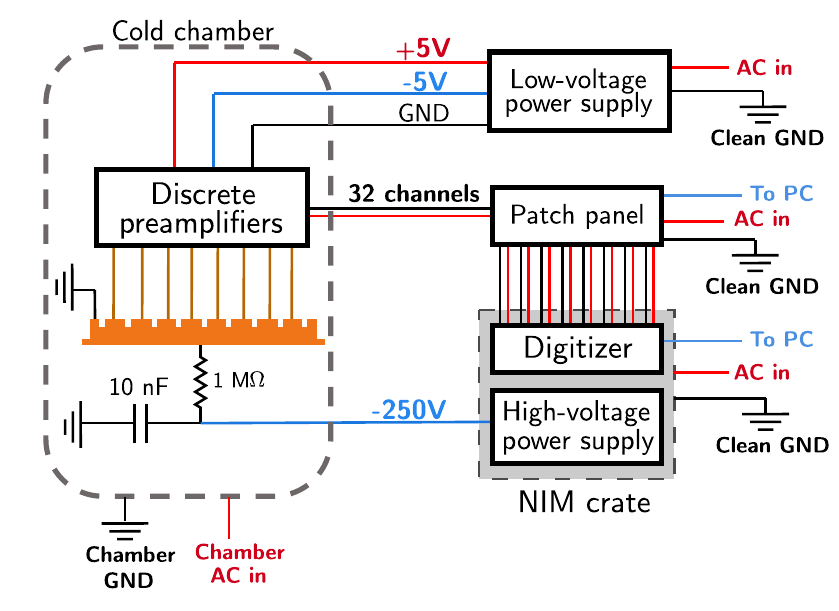}
    \caption{Electrical connection diagram for the Si(Li) module testing system. For clarity, only one of the four Si(Li) detectors is shown, with all four detectors receiving the same --250-V bias via the 1-M$\Omega$ resistor. The electrical ground for the Si(Li) module and preamplifier holder boards is defined by the LVPS. The exterior electronics (LVPS, patch panel, and NIM crate) receive clean AC power and grounding from the isolation transformer, with the lab computer (PC) and cold chamber receiving AC power and grounding from the building mains. For more details see Sec.~\ref{sec:testing_system}.}
    \label{fig:electronic_diagram}
\end{figure}

 As the FEBs and ASICs were being developed simultaneously with the Si(Li) detector testing, we require a different method of detector readout for the lab testing described in this work. To that end, we replace the FEB and top polypropylne window with a single FR4 board. Each board contains a 1-M$\Omega$ resistor for overcurrent protection and a 10-nF capacitor for noise protection as shown in Fig.~\ref{fig:electronic_diagram}. Spring-loaded conductive pins (``pogo pins'') mounted to the underside of the window couple the Si(Li) strips to conductive pads on the top surface of the window for signal readout via charge-sensitive discrete preamplifiers.

\subsection{Custom Si(Li) Module Testing System}
The GAPS module testing systems at the Massachusetts Institute of Technology (MIT) and the University of Hawai'i at M{\=a}noa (UHM) use largely identical systems and commercially-available components. The purpose of the tests was to characterize the detector energy resolutions based on known x-ray lines with energies in the range of the exotic-atom transition x-rays that GAPS is uniquely designed to detect (${\sim}$20--100 keV). Each testing setup itself is composed of several major subsystems: the environmental chamber, the cable feedthrough system, the exterior electronics, the discrete preamplifiers, and the radioactive source.
\subsubsection{Environmental Chamber}
We use SUN Electronics EC13 chambers to provide a temperature- and humidity-controlled environment during testing, as well as to facilitate the electrical connections to the module. The GAPS Si(Li) detectors are expected to operate at a temperatures lower than approximately --40$\,^{\circ}$C; therefore, we test detectors at a conservative --37$\,^{\circ}$C. The chambers include inlets for liquid and gaseous N$_2$ for cooling and humidity suppression, respectively, as well as a built-in temperature display and interior fan. The SUN chamber doors were modified in-house to provide holes for cable feedthrough (discussed later in this section) and detector module mounting to the attached tray. 
\subsubsection{Discrete Preamplifiers}
We use custom charge-sensitive discrete preamplifiers to extract signals from the Si(Li) strips and perform initial pulse shaping. The preamplifiers are based on Ref.~\cite{Fabris:1999}, with each preamplifier including a 100-M$\Omega$ feedback resistor, a 0.5-pF feedback capacitor, and a low-noise JFET with capacitance ${\sim}$10 pF. The preamplifiers have a noise floor of ${\sim}$2 keV FWHM. Each Si(Li) strip is connected to a single preamplifier, with eight preamplifiers (i.e., the readout for each Si(Li) detector) mounted into a holder board. The holder boards are mounted to the FR4 top window of the module with conductive screws and standoffs, providing both mechanical and electrical (i.e., ground) connections.  Finally, to shield the preamplifiers and Si(Li) detectors from external noise, we create custom Faraday cages from 0.6-mm-thick aluminum sheets, folded into five-sided rectangular prisms with dimensions ${\sim}10\text{\,cm}\times 10\text{\,cm}\times 10\text{\,cm}$ to place atop each detector/holder board, with small holes cut into each to accommodate the signal and low-voltage power cables.
\subsubsection{Cable Feedthrough System}
The cable feedthrough systems provide connections for the low-voltage discrete preamplifiers, the high-voltage detector bias, the individual discrete preampflifier signal channels (one per detector strip), and various backup cables. They are composed of circular non-conductive black Teflon flanges custom-machined for this purpose, with high-fidelity RG316 coaxial cables potted through the flanges with the use of Torr Seal. The coaxial cables had their exterior plastic insulation stripped at a small ($\sim$1-2 mm) length so they could be to be bonded to the Teflon flange with cyanoacrylate adhesive prior to the final application of Torr seal. Circular openings consistent with the flange dimensions were machined into the cold chamber doors, and the flanges screwed into place with a thin O-ring and potted with Torr Seal to minimize heat leak during operation.
\begin{figure}[t]
    \centering
    
    \label{fig:my_label}
\end{figure}

\subsubsection{Power and Data Acquisition}
We use several commercial electronic modules in conjunction with the testing chamber. A CAEN 1471A high-voltage power supply (HVPS) is used to provide the --250V reverse bias necessary to fully deplete the detectors, as well as to monitor the supply current. A \red{14-bit} CAEN N6725 digitizer is used to sample the voltage waveforms \red{at 250 MS/s} from eight Si(Li) strips simultaneously. Both CAEN modules are mounted into a NIM crate whose onboard cooling fan was disabled and replaced with an external fan to reduce electromagnetic noise pickup. The discrete preamplifiers are supplied with $\pm 5\text{\,V}$ and ground connections via a commercial HY3005F-3 low-voltage DC power supply (LVPS). The 32 preamplifier signal outputs (one per Si(Li) strip) are connected to a CXAR-64/MF patch panel, from which up to eight inputs can be selected and sent to the digitizer (the remaining inputs are terminated by 50$\,\Omega$). The patch panel and digitizer are both controlled by a computer running custom software. Finally, we use a Tripp-Lite 1-kW power isolation transformer to provide steady, low-noise AC power and grounding to the NIM crate, LVPS, and patch panel (the computer and cold chamber receive power directly from the building mains).
\subsubsection{Radioactive Source} 
We use a sealed 1-mCi $^{109}\text{Cd}$ source to study the energy response of the GAPS detectors. This isotope has several properties that make it ideal for GAPS detector testing~\cite{KUMAR20161}. First, its ${\sim}$88-keV decay $\gamma$-ray lies in the energy range of the exotic-atom x-rays that GAPS is designed to detect. Second, $^{109}\text{Cd}$ decays by electron capture to the first excited state of $^{109}\text{Ag}$, and does not produce \red{high-energy} $\alpha$- or $\beta$-particles that might lead to unwanted detector backgrounds during testing. \red{(We note that $^{109}\text{Cd}$ emits Auger and conversion electrons with kinetic energies ranging from ${\sim}2.6\text{\,keV}$ to ${\sim}87.9\text{\,keV}$, though these electrons are easily stopped by the sealed source, aluminum Faraday cages, and FR4 board above the detectors~\cite{estar}.)} Finally, the decays of $^{109}\text{Cd}$ also produce fluorescence x-rays with energies between ${\sim}$22--25 keV, allowing the study of low-energy detector behavior, \red{though this is outside the scope of this analysis}.

\section{{\label{sec:testing_flow}}Detector Processing and Testing}

\subsection{Detector Pre-selection Model}
To enable efficient testing and integration of the GAPS Si(Li) detector modules, we developed a pre-selection model to group detectors of similar quality based on their room-temperature leakage current at --250\,V bias. The pre-selection criteria are shown in Table~\ref{tab:test_results}, and are based on several requirements. First, the power system developed to supply high voltage to the flight instrument imposes a current limit of ${\sim}6\,\upmu\text{A}$ per detector module at --250\,V (conservatively, ${\sim}1\,\upmu\text{A}$ per detector). Second, the ASIC readout developed for GAPS places an upper limit of 50 nA on each strip's leakage current, above which the strip cannot be read out. Initial testing of $\mathcal{O}(20)$ bare (i.e., non-passivated) detectors showed first that detectors with total room-temperature leakage current below 300$\,\upmu\text{A}$ remained below the required $1\,\upmu\text{A}$ at --37$\,^\circ\text{C}$ (i.e., were functional detectors); second, that strips with room-temperature leakage current below 12.5$\,\upmu\text{A}$ remained below the required 50 nA at --37$\,^\circ\text{C}$ (i.e., were functional strips); and third, that strips with room-temperature leakage current below $2\,\upmu\text{A}$ had ASIC-projected energy resolution below 4 keV FWHM at --37$\,^\circ\text{C}$ (i.e., were suitable for x-ray spectroscopy). We note that all detectors were assembled into modules and tested at --37$\,^\circ\text{C}$, even if the pre-selection criteria indicated they might not be operable.
\input{sili_table}

\subsection{Detector Storage and Shipping}
Our main goal during detector storage and shipping is to provide as inert an environment for the detectors as possible. Following passivation at Columbia University (CU), detectors are returned to their plastic cases and placed into antistatic bags. Each antistatic bag contains a silica desiccant pack and is vacuum-sealed before being placed into a freezer maintained at --20$\,^\circ\text{C}$ to suppress the rate of lithium drift. When detectors are selected for testing, they are removed from the freezer and (still sealed in their antistatic bags) allowed to slowly return to room temperature in a dry environment before being assembled into modules. Each module (including FR4 top window, four detectors, and the aluminized polypropylene bottom window) is placed into an antistatic bag with silica desiccant and lightly vacuum-sealed, and subsequently placed into another antistatic bag and lightly vacuum-sealed. The module bags are then enclosed in custom antistatic foam inserts before being packaged for shipment to the testing sites (MIT and UHM). At the testing sites, the modules (still in their antistatic bags) are placed into room-temperature dryboxes maintained at relative humidity $\lesssim$15\%. The modules typically remain at the testing sites for one to two weeks before being tested and subsequently packaged for return to CU for long-term storage. We have also tested several detector modules after nearly two years in storage, both in the room-temperature dryboxes and in the freezers, and observed no significant change in detector performance.
\subsection{Module Testing Protocol}
When a module is to be tested, we remove it from drybox and affix the aluminum Z-bars to the four corners of each module to elevate it ${\sim}$2 cm above the tray. The module is then mounted to the SUN chamber tray with non-conductive PEEK screws, and the four preamplifier holder boards are mounted and secured to the top FR4 window. Next, we connect the signal readout and low-voltage DC power cables to each preamplifier holder board, taking care to ensure that the readout cables are connected to the high-gain output. Finally, we secure the aluminum Faraday cage above each detector/holder board with conductive tape, ensuring that each cage is grounded to the aluminum frame of the module. Immediately before mounting the door assembly to the SUN chamber, we place a sealed 1-mCi $^{109}\text{Cd}$ source atop the Faraday cages, in the center of the module, to provide a consistent x-ray flux across the detectors.
\par With the module now sealed in the SUN chamber, we cool it to --37$\,^\circ\text{C}$ over approximately 30 minutes. We also supply dry nitrogen gas at a rate of a few liters per minute to further suppress any residual humidity within the chamber. When the chamber temperature readout indicates the set temperature has been reached, we begin to bias the detectors at a rate of 3 V/s until --250\,V is reached. We find that an additional ${\sim}$15 minutes are required for the high-voltage current to decrease to its equilibrium value (typically below 100 nA for high-quality detectors). The data-acquisition software is described in the next section; here, we note that three minutes is sufficient to acquire several thousand events per strip for a single detector. During acquisition, we turn off the chamber interior fan, heating, and cooling systems to suppress noise (i.e., only the temperature readout remains active). Following the three-minute period, these systems are turned back on and the chamber is allowed to re-cool for at least ten minutes.
\par Once x-ray events have been acquired for all detectors, we ramp the detector bias down to zero at a rate of 3\,V/s and allow the chamber to return to room temperature over approximately 30 minutes. The nitrogen gas flushing is particularly important for humidity suppression during the warm-up, as the liquid nitrogen valve is closed. Once the chamber has returned to room temperature, we remove the door assembly and disconnect the module's mechanical and electrical connections. The module is then returned to the drybox for short-term storage.
\subsection{Data Acquisition and Processing}
For each Si(Li) detector strip, we must acquire and fit x-ray energy spectra using at least ten peaking times $\tau_\text{p}$, ranging from 0.5--30$\,\upmu\text{s}$, to fully sample the noise model described in Sec.~II-D. (The pulse peaking time $\tau_\text{p}$ is the time required for the shaped pulse to reach its maximum~\cite{Grupen:2008zz}.) With more than 1000 detectors to test, any method involving the acquisition of spectra at individual peaking times would require an impractical amount of time.    
\par To increase the module testing speed, we implement custom software to read each strip's voltage waveforms directly from the N6725 digitizer for later offline processing. To fully sample the preamplifier signal, which has a decay time constant $\tau_\text{d} \sim 50\,\upmu\text{s}$, we record 160\,$\upmu\text{s}$ per waveform, of which $4\,\upmu\text{s}$ is pre-trigger to establish a baseline. Each waveform is sampled at a rate of 250 MS/s, and the dynamic range of the digitizer is set to 0.5V peak-to-peak with 80\% DC offset. We acquire waveforms from eight strips (i.e., one detector) simultaneously for three minutes, producing several thousand events per strip depending on proximity to the x-ray source. The waveforms for each detector are saved in a \textsc{root} file for offline analysis.
\par
\begin{figure}
    \centering

    \hspace{-0.55cm}
    \subfloat{
        \includegraphics[width=0.95\columnwidth]{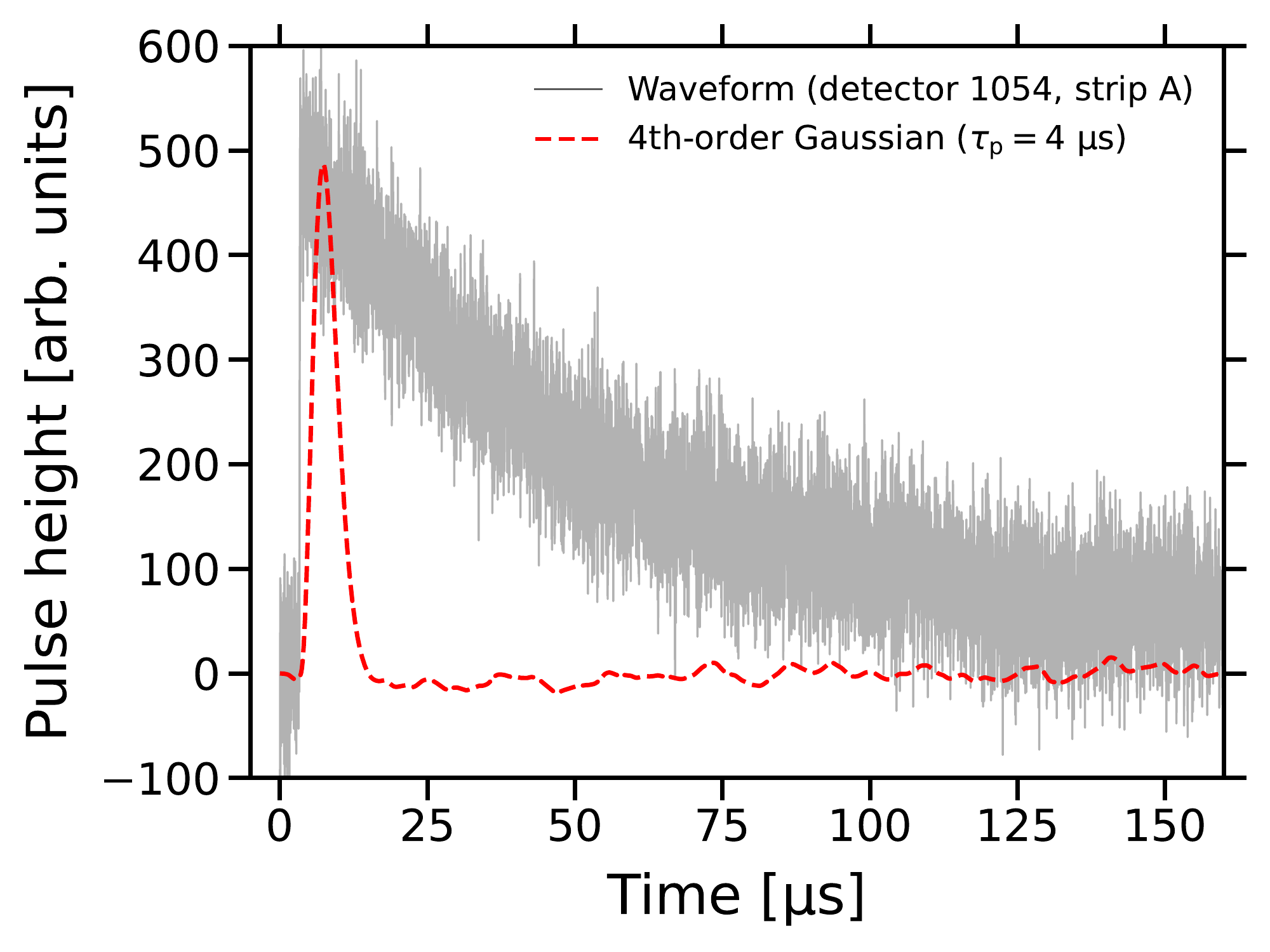}
    }

    
    \subfloat{
        \includegraphics[width=0.95\columnwidth]{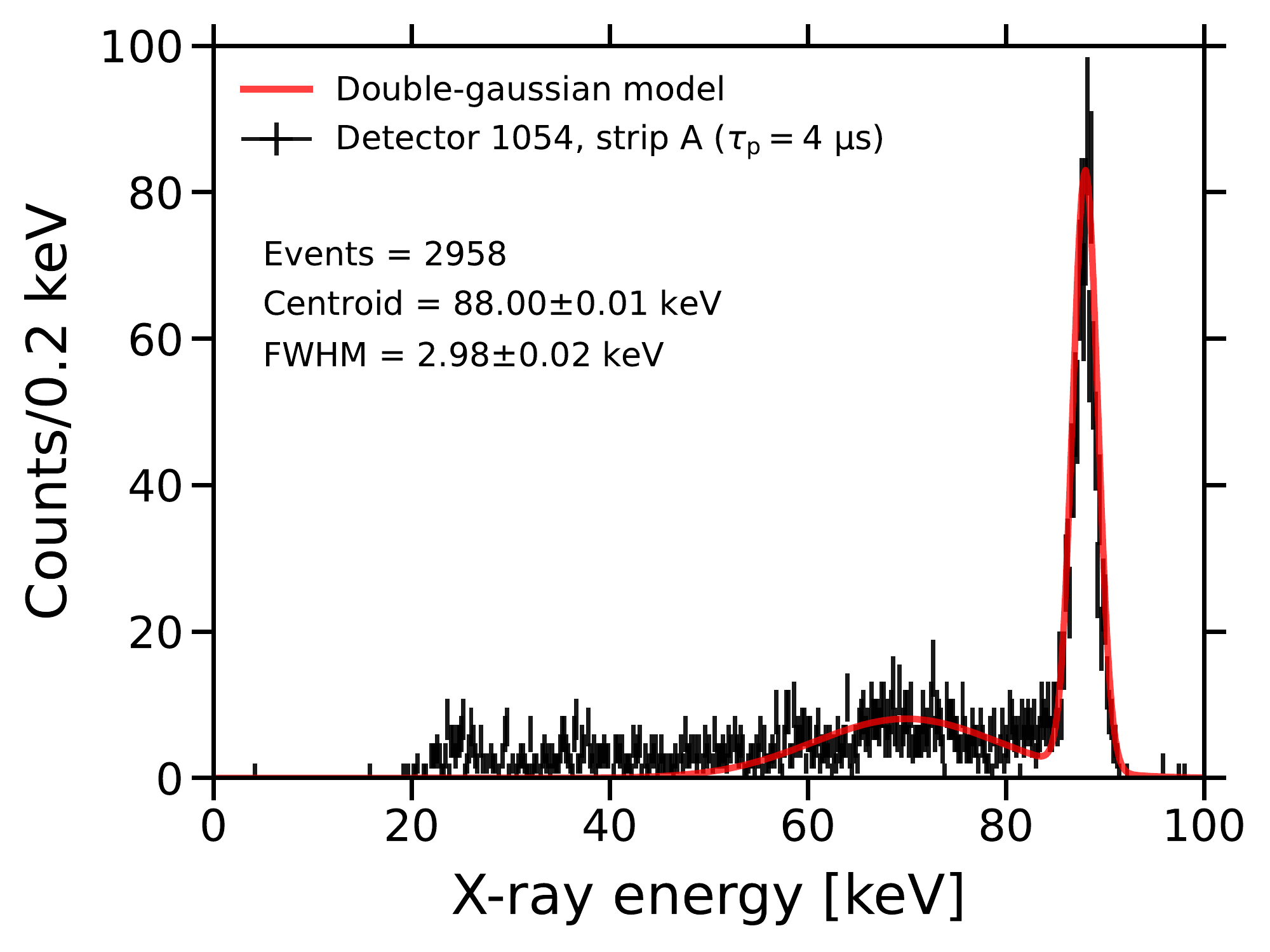}
    }
    
    \vspace{0.25cm}

        \subfloat{
        \includegraphics[width=0.95\columnwidth]{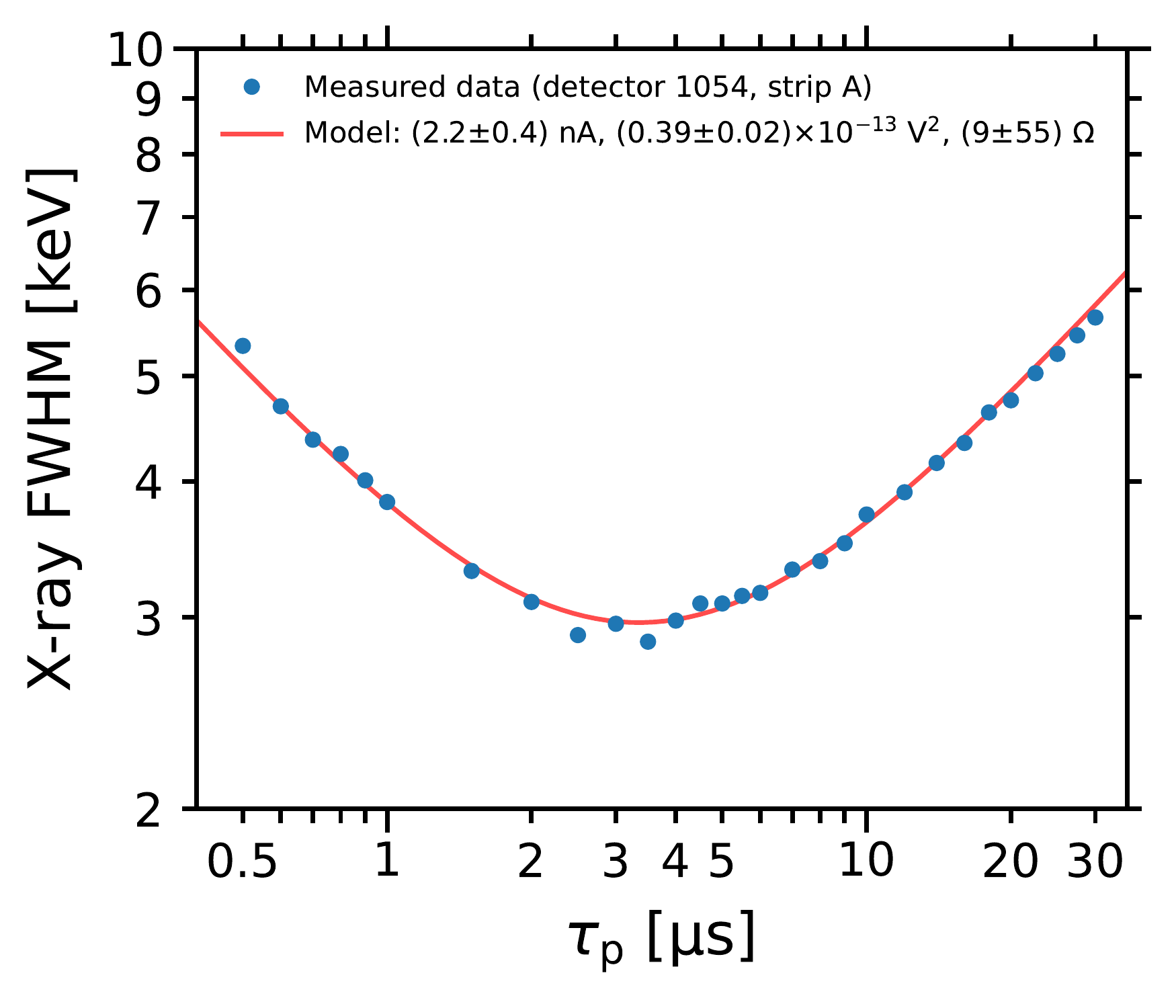}

    }

    \caption{The data-analysis workflow for strip A of GAPS Si(Li) detector 1054 tested at --37\,$^\circ\text{C}$. \textbf{Top}: a single baseline-subtracted waveform and the corresponding $\tau_\text{p} = 4\,\upmu$s Gaussian-shaped pulse. \textbf{Middle}: the corresponding $^{109}\text{Cd}$ x-ray energy spectrum, binned to 0.2 keV/bin for presentation. \textbf{Bottom}: the best-fit x-ray noise model from Eq. (1) as a function of $\tau_\text{p}$. FWHM uncertainties are comparable to the size of the data points.}
    \label{fig:workflow}
\end{figure}

 To construct x-ray spectra for each strip, we first shape the waveforms with fourth-order Gaussian filters~\cite{Ohkawa:1976} implemented in software, \red{as shown in Fig.~\ref{fig:workflow}}. We use a set of peaking times $\tau_\text{p}$ ranging from  0.5--30$\,\upmu\text{s}$, allowing us to sample the three terms in the noise model described in Sec.~\ref{sec:analysis_results}. The amplitude of the shaped pulse is proportional to the x-ray energy deposition. For each detector strip, we generate one spectrum per $\tau_\text{p}$ value. We fit the x-ray spectra with two Gaussian functions representing the $^{109}\text{Cd}$ full-absorption peak and Compton shoulder, with both Gaussians' peak location, width (FWHM), and amplitude allowed to vary. Finally, we convert these spectra to energy by normalizing the best-fit peak centroid to the known ${\sim}$88-keV full-absorption peak from $^{109}\text{Cd}$. Previous testing with the 59.5-keV and 88-keV $\gamma$-rays from $^{241}\text{Am}$ and $^{109}\text{Cd}$, respectively, has demonstrated the linearity of the GAPS Si(Li) energy response in the range ${\lesssim}100$ keV relevant for this work~\cite{Rogers:2019avj}.
\par Crucially, a single three-minute data acquisition per detector is sufficient to construct a noise model for all eight strips. Including re-cooling after each detector is tested, this method allows us to acquire a full set of data for each module in less than one hour; thus, our throughput is mainly limited by the ${\sim}$two hours required to cool the chamber, warm it back up to room temperature, and mount/unmount the detector module from the chamber.

\section{\label{sec:analysis_results}{Analysis and Results}}
There were three primary goals of the Si(Li) detector testing effort. First, we sought to demonstrate that the detectors' noise characteristics were consistent with the noise model described in Eq. (1). Second, we used the measured x-ray energy resolution to sort the detectors' quality and inform their placement within the payload. Finally, we used the extracted noise-model parameters to model the detectors' expected performance in the flight instrument.

\subsection{\label{sec:noisemodel}Noise Model}
For each strip, we fit the full-width-at-half-maximum (FWHM) energy resolution at each peaking time $\tau_\text{p}$ to the following noise model appropriate for the discrete preamplifiers \cite{Rogers:2019avj}:

\begin{align}{\label{eqn:preamp_noisemodel}}
    \text{ENC}^2 &= \left(2qI_\text{leak}+\frac{4kT}{R_p}\right)F_i\tau_\text{p} \\
    &\;\;\;\;+4kTC_\text{tot}^2\left(R_s+\frac{\Gamma}{g_m}\right)\frac{F_\nu}{\tau_\text{p}}+2\pi A_f C_\text{tot}^2 F_{\nu f} \nonumber
\end{align}
Here, ENC is the equivalent noise charge, related to the detector FWHM energy resolution via 
\begin{equation}
    \text{FWHM} \simeq 2.35\epsilon\frac{\text{ENC}}{q},
\end{equation}
where ${\epsilon \approx 3.6\text{\,eV}}$ is the ionization energy per electron-hole pair in silicon~\cite{Klein:1968a} and $q$ is the elementary charge. \red{The noise model for each strip contains three free parameters: the strip leakage current $I_\text{leak}$, the coefficient of $1/f$ noise $A_f$, and the series resistance $R_s$ arising mainly from the pressure-mounted pogo-pin readout (though as described below, we are principally interested in $I_\text{leak}$ and $A_f$). The other terms are as follows:} $C_\text{tot}$ is the total capacitance to be defined shortly, $k$ is the Boltzmann constant, $T$ is the absolute temperature, $\Gamma = 1$ is \red{calculated from the geometry of the preamplifier JFET}, $g_m = 0.1\text{\,mS}$ is the room-temperature transconductance of the input FET, and $R_p = 100\text{\,M}\Omega$ is the parallel resistance in the preamplifier. \red{(We note that any deviation from the assumed value $\Gamma = 1$ adopted in Ref.~\cite{Rogers:2019avj} and in this work would lead to a small variance in the best-fit value of $R_s$; however, $R_s$ is not used in the ASIC noise projection.)} The form factors $F_i = 0.45$, $F_\nu = 1.02$, and $F_{\nu f} = 0.52$ are appropriate for the fourth-order Gaussian shaper. The total input capacitance is given by $C_\text{tot} = C_\text{strip} + C_\text{FET} + C_\text{int} + C_\text{stray}$, including contributions from the Si(Li) strip ($C_\text{strip} \approx 37 \text{--39\,pF}$ at --250\,V), the preamplifier FET ($C_\text{FET} \approx 10 \text{\,pF}$), and any inter-electrode ($C_\text{int}$) and stray ($C_\text{stray}$) capacitance. Previous work \cite{Rogers:2019avj} has shown that $C_\text{int} + C_\text{stray} \approx 20\text{\,pF}$, so we fix $C_\text{tot} = 70\text{\,pF}$ during our noise-model fits to \red{lift} the degeneracy between $C_\text{tot}$, $A_f$, and $R_s$. Lastly, we note that the minimum achievable resolution for strips with negligible $I_\text{leak}$, $A_f$, and $R_s$ is ${\sim}$2 keV FWHM, owing to thermal noise in the 100-M$\Omega$ resistor and the transconductance of the FET.
\par The GAPS flight instrument will use a custom ASIC rather than discrete preamplifiers to read out the Si(Li) detectors. As fabrication and testing of the ASIC was ongoing during the Si(Li) detector testing campaign, it was necessary to develop a noise model to predict the detector resolution expected with the flight ASIC readout. The ASIC noise model is \cite{Manghisoni:2021}
\begin{align}
    \text{ENC}^2 &= 2q(I_\text{leak} + I_{k} )F_i \tau_\text{p} + 4kT R_\text{eq}C_\text{eq}^2\frac{F_\nu}{\tau_\text{p}} \\
    &\;\;\;+2\pi A_f C_\text{eq}^2 F_{\nu f}, \nonumber
\end{align}
where $A_f$ is either $0.87\times 10^{-13}\,\text{V}^2$ (derived from simulations) or the fitted value from the discrete preamplifier testing, whichever is greater; $I_\text{leak}$ is the strip leakage current; $I_k \approx 2.5\text{\,nA}$ is the effective current in the analog channel; ${C_\text{eq} = C_\text{strip} + C_\text{amp}}$ is the equivalent parallel capacitance of the strip (${C_\text{strip} \approx 37\text{\,pF}}$) and amplifier input ($C_\text{amp} \approx 5\text{\,pF}$); $R_\text{eq}\approx 40\,\Omega$ is the equivalent \red{resistance of the wirebond and input PCB trace (taking the place of $R_s$)}; and $F_i = 0.64$, $F_\nu =0.85$, and ${F_{\nu f} = 0.54}$ are the form factors of the unipolar semi-Gaussian $\text{(RC)--(CR)}^2$ shaper. \red{With these parameters, the ASIC readout generally achieves the best energy resolution for $\tau_\text{p}\sim 1\,\upmu\text{s}$, somewhat lower than the ${\sim}4\,\upmu\text{s}$ achieved with the discrete preamplifiers. To minimize the contributions from leakage current at longer peaking times, the ASIC operates in the range 0.25--1.8$\,\upmu\text{s}$.} Tests are ongoing to validate this noise model using the final flight components (ASIC, front-end board, etc), so this work will use a conservative value $C_\text{eq} = 70\text{\,pF}$ to account for any un-modeled stray capacitance.

\subsection{Detector Quality}
We categorize the quality of each Si(Li) strip and detector into three broad groups based on their projected ASIC resolution, as shown in Table~\ref{tab:test_results}. The flight ASIC readout and high-voltage power system place limits on the allowed leakage current of 50 nA/strip and 1 $\upmu$A/detector, respectively, which we take as criteria for \textit{non-usable} strips and detectors at --37$\,^\circ\text{C}$. Also included in these categories are strips which do not produce stable voltage waveforms during testing and detectors with $\ge$4 non-usable strips. As shown in Table~\ref{tab:test_results}, the yield rate of non-usable strips and detectors is ${\sim}$10\%, though this is likely an overestimate, as more mundane issues (e.g., loose cable connections) can mimic the appearance of non-functional strips.
\par For the functional strips and detectors, the key performance metric is the ASIC-projected x-ray energy resolution. To distinguish between the x-rays from antiprotonic and antideuteronic exotic atoms, we require the projected ASIC energy resolution determined using Eq. (4) to be $\leq$4 keV FWHM at -37$\,^\circ\text{C}$. Strips which meet this criterion are designated \textit{x-ray-quality strips}, and detectors with $\ge$7 x-ray strips are designated \textit{x-ray-quality detectors}. Strips whose projected ASIC resolution is $>$4 keV FWHM are designated \textit{tracking-quality strips}, as they are still able to provide position and energy-deposition information for charged particles.
\par
\begin{figure}[t]
\centering
\includegraphics[width=0.95\columnwidth]{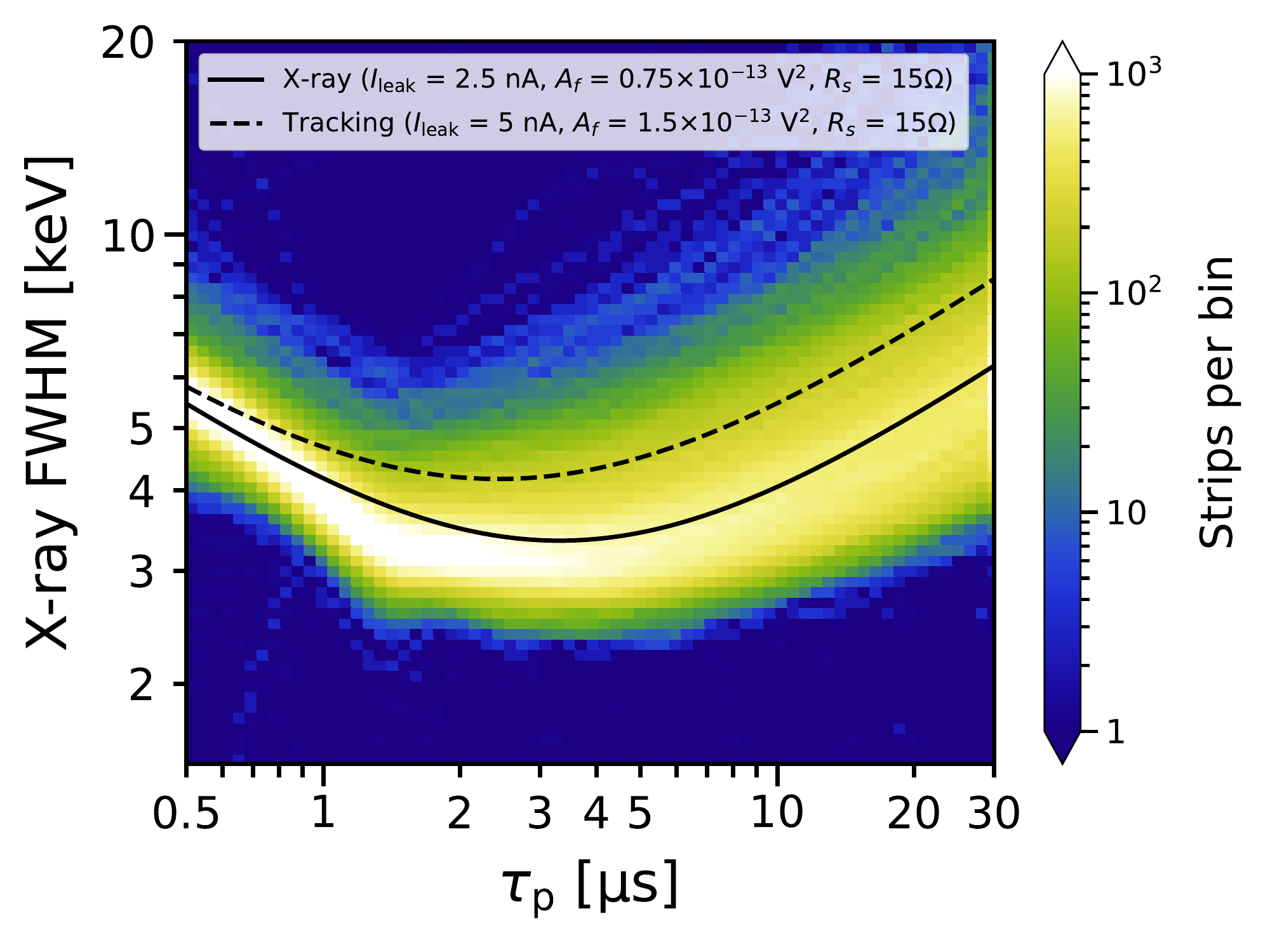}
\caption{Heatmap showing Si(Li) strip energy resolution obtained with the discrete preamplifiers as a function of $\tau_\text{p}$. The data have been interpolated with a polynomial for presentation purposes only, and the solid and dashed black lines indicate representative noise models for x-ray and tracking strips, respectively.}  
\label{figure:noisemodel_all_strips}
\end{figure}

\begin{figure}[t]
\includegraphics[width=0.95\columnwidth]{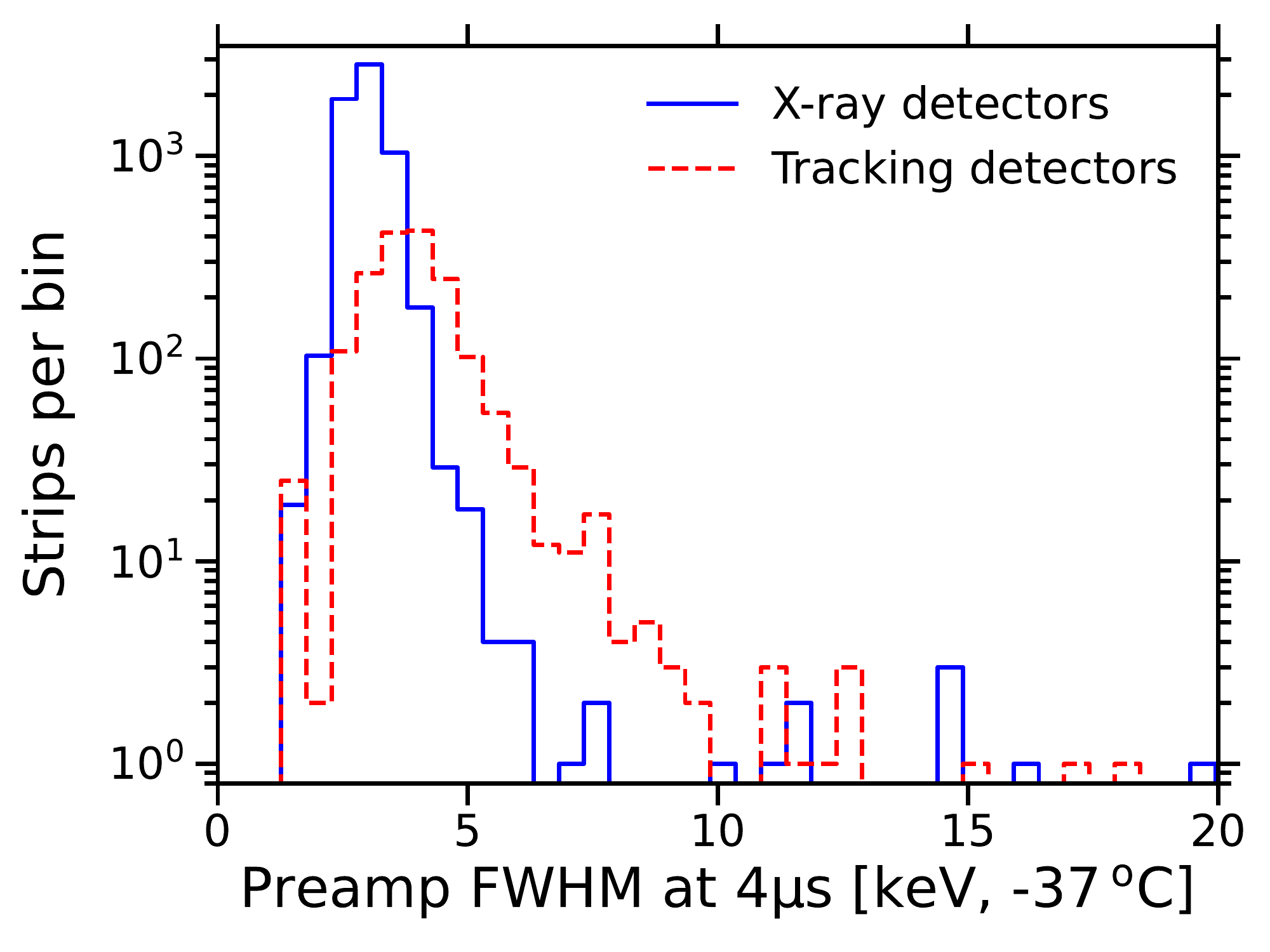}
\caption{Histogram of Si(Li) strip x-ray energy resolutions measured with discrete preamplifiers at 4$\,\upmu\text{s}$ and --37$\,^\circ\text{C}$ for x-ray detectors (solid blue) and tracking detectors (dashed red).}  
\label{fig:strip_fwhm_histogram}
\end{figure}

 In Fig.~\ref{fig:strip_fwhm_histogram} we show the distribution of strip x-ray energy resolutions obtained with discrete preamplifiers at ${\tau_\text{p} = 4\,\upmu\text{s}}$, separated into x-ray and tracking-quality detectors. (As shown in Fig.~\ref{figure:noisemodel_all_strips}, strips with typical values of $I_\text{leak}$ and $A_f$ coupled to discrete preamplifiers attain their optimal energy resolution near this value of $\tau_\text{p}$.) We note several features of the distributions. First, the most-probable value for the energy resolution of x-ray-quality strips is ${\sim}$3 keV FWHM, but there is a tail extending out to ${\sim}$8 keV, with some strips exceeding 10 keV. The distribution of strip resolution for the tracking-quality detectors peaks at ${\sim}$4 keV FWHM and has a longer tail, though a significant number of strips on tracking-quality detectors have FWHM $<4$ keV. This reflects our observation that many detectors have ${\sim}$6--7 x-ray-quality strips and ${\sim}$1--2 tracking-quality strips, indicating that poorly-performing (or even non-functional) strips do not necessarily affect their neighbors within a detector.
\par
\begin{figure}[h]
\centering
\includegraphics[width=0.975\columnwidth]{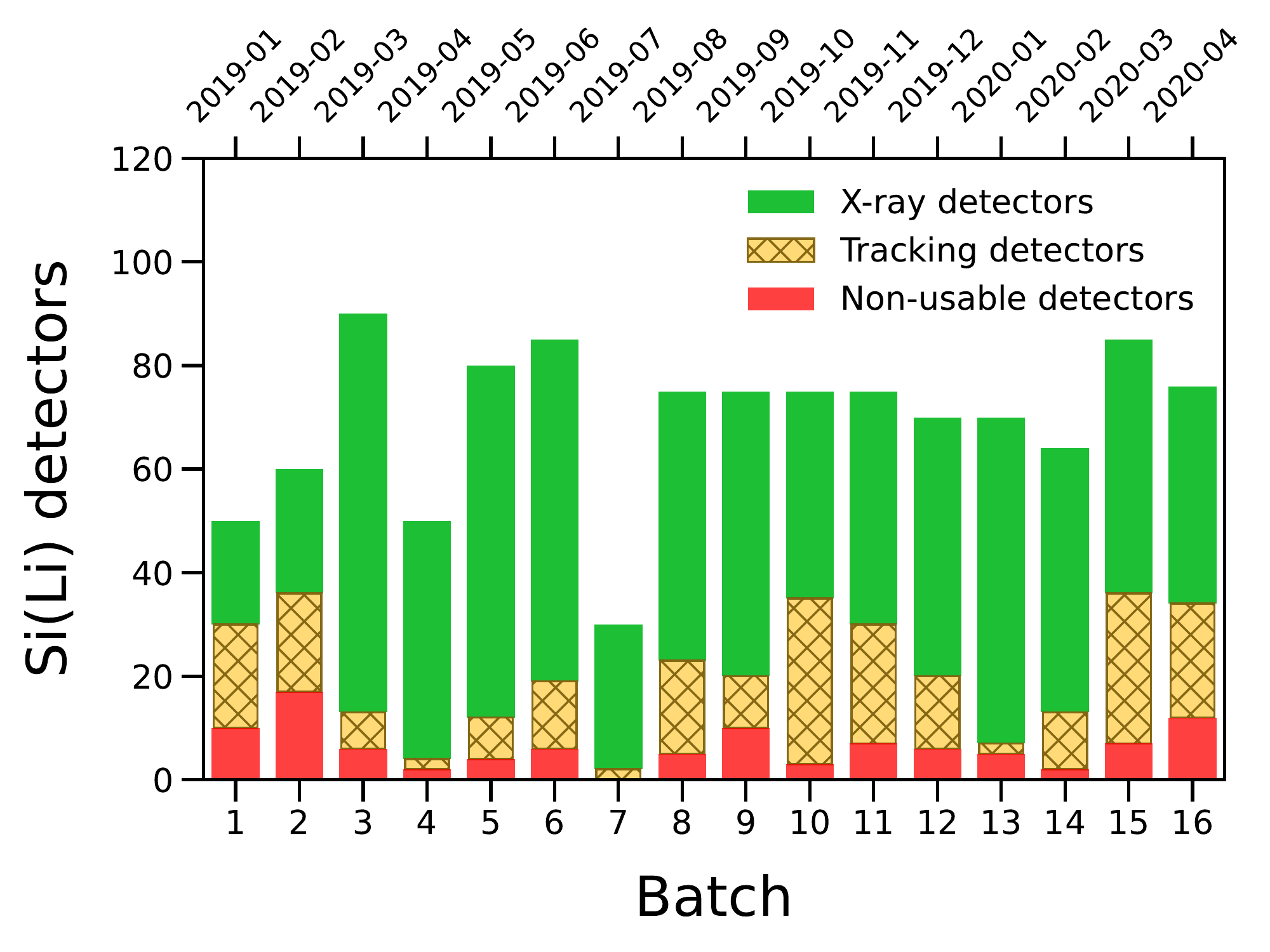}
\caption{Stacked bar chart showing the number of x-ray (green, top), tracking (hatched yellow, middle), and non-usable (red, bottom) detectors from each production batch. The approximate \red{production} date ranges of each batch are also shown. \red{The decrease in the number of detectors produced in batches 1, 2, 4, and 
 7 corresponds to Japanese public holidays during those months~\cite{Kozai:2021apo}.}}  
\label{fig:batch_quality}
\end{figure}

 Finally, we examine for the first time the yield of x-ray, tracking, and non-usable detectors across the entire fabrication timescale (Jan. 2019 to Apr. 2020). As described in \cite{Kozai:2019xlp,Kozai:2021apo}, the fabrication was divided into 16 sequential batches, each lasting approximately one month. The results are shown in Fig.~\ref{fig:batch_quality}. We note that batches 1 and 2 have a noticeably higher rate of tracking and non-usable detectors than the following batches. Following batch 2, the lithium-drifting procedure was modified to terminate either at ${\sim}6200$ minutes or when the drift current began to increase exponentially above ${\sim}$15--25 mA, whichever occurred first. This allowed us to better control the depth of the drifted region, and hence the detector leakage current and capacitance~\cite{Kozai:2019xlp,Kozai:2021apo}. These results further demonstrate that large-area Si(Li) detectors with good x-ray resolution can be mass-produced.

\subsection{Detector Noise Parameters}
\begin{figure}[t]
    \centering
    
    \subfloat{
        \includegraphics[width=0.975\columnwidth]{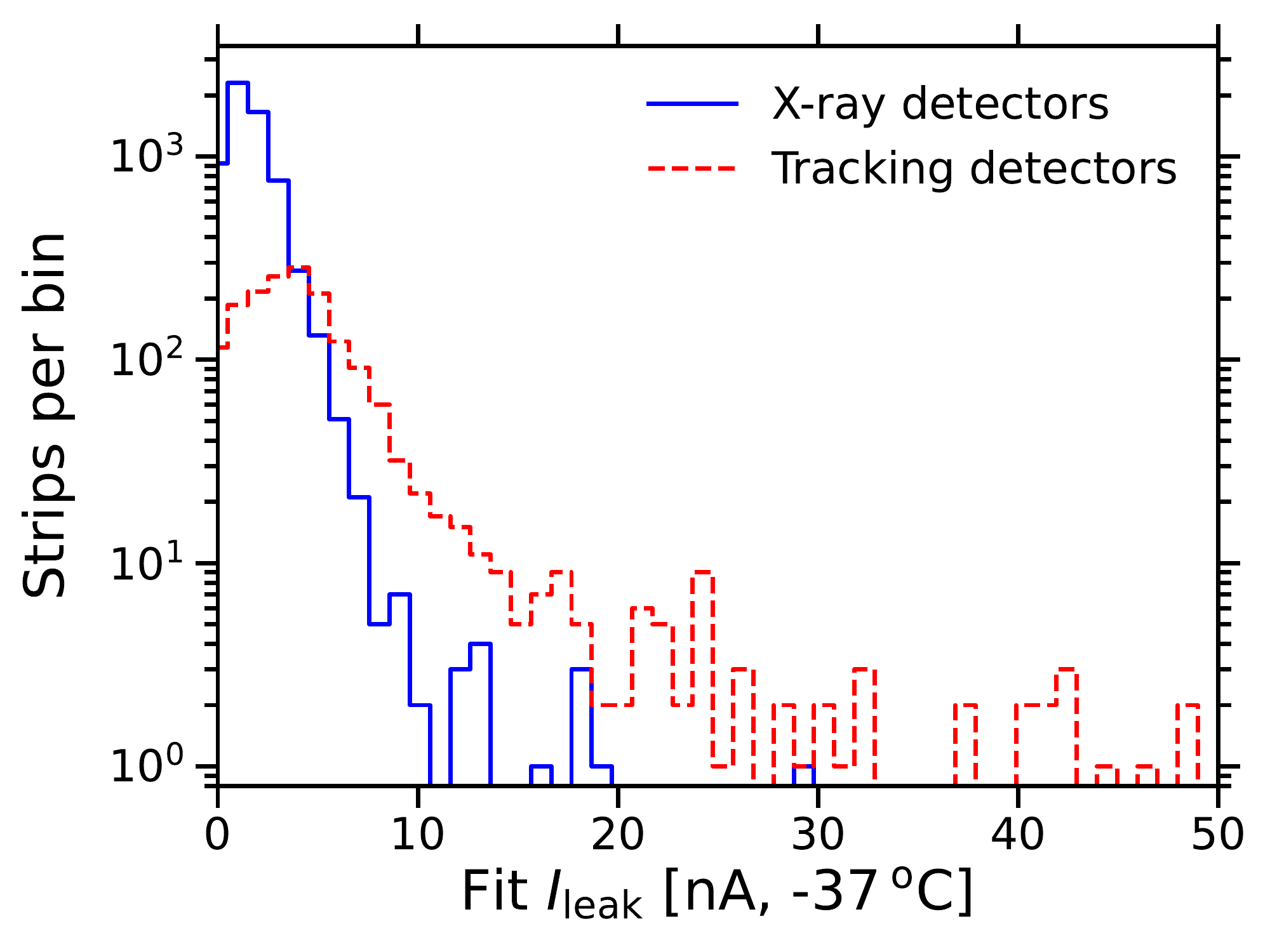}
    }
    
    \subfloat{
        \includegraphics[width=0.95\columnwidth]{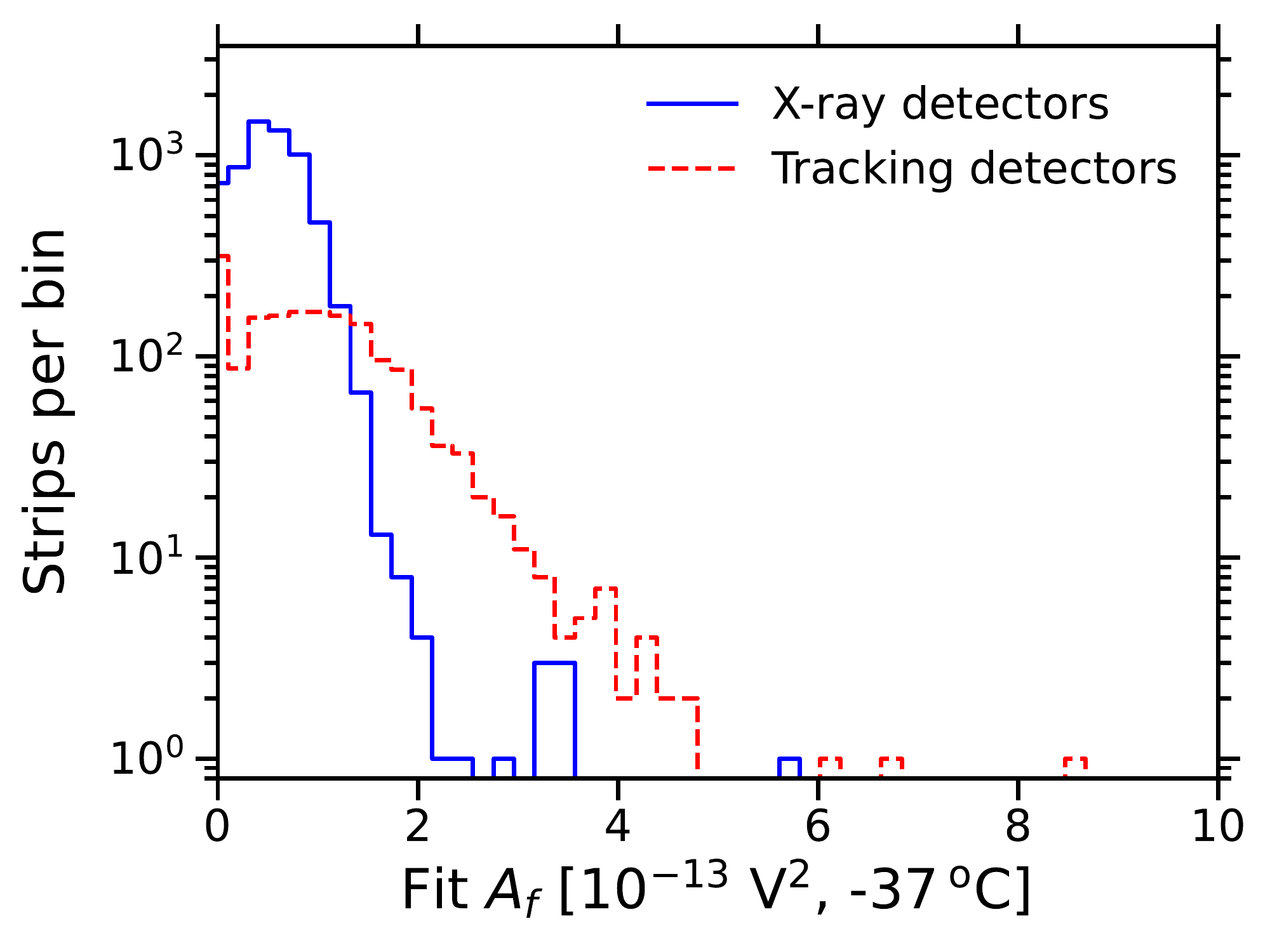}
    }

    \caption{Histograms showing fitted $I_\text{leak}$ (top) and $A_f$ (bottom) for Si(Li) strips on x-ray detectors (solid blue) and tracking detectors (dashed red). Values of $I_\text{leak}$ and $A_f$ less than zero are placed in the leftmost bin. There are ${\sim}$30 strips with fit $I_\text{leak} > 50\text{\,nA}$ and ${\sim}20$ strips with $A_f > 10^{-12}\text{\,V}^2$, though since the discrete preamplifiers are expected to saturate near 50 nA, those results should be taken with care.} 
    \label{fig:ileak_af_hist}
\end{figure}

As described in Sec. II-E, the strip leakage current $I_\text{leak}$ and $1/f$ coefficient $A_f$ are the main drivers of the x-ray energy resolution at the $\mathcal{O}(\upmu\text{s})$ shaping timescales of interest for GAPS. Figure \ref{fig:ileak_af_hist} shows the distributions of $I_\text{leak}$ and $A_f$ for the x-ray and tracking-quality detectors. We note several features. First, the most-probable values of strip $I_\text{leak}$ for x-ray and tracking-quality detectors are ${\sim}$2 nA and ${\sim}$4 nA, respectively, with the tracking-quality detectors having a significantly longer tail. We also note that ${\sim}120$ strips (i.e., less than 1.5\% of the total) have best-fit values of $I_\text{leak}$ and/or $A_f$ that are consistent with zero or even slightly negative. This occurs most often when the other noise parameters are small (e.g., $I_\text{leak} \lesssim 0.5\text{\,nA}$ or ${A_f \lesssim 0.2\times 10^{-13}\text{\,V}^2}$), since the internal preamplifier components dominate the fit for these especially low-noise strips. Furthermore, these low-noise strips are also sensitive to deviations in $C_\text{tot}$ away from the assumed 70 pF value, though as discussed in Sec.~\ref{sec:noisemodel} we needed to freeze $C_\text{tot}$ to break the degeneracy with $A_f$. When calculating the projected ASIC energy resolution using Eq. (3), we do not allow $I_\text{leak}$ or $A_f$ to go below zero. 
\par
\begin{figure}
    \centering
    
    \subfloat{
        \includegraphics[width=0.95\columnwidth]{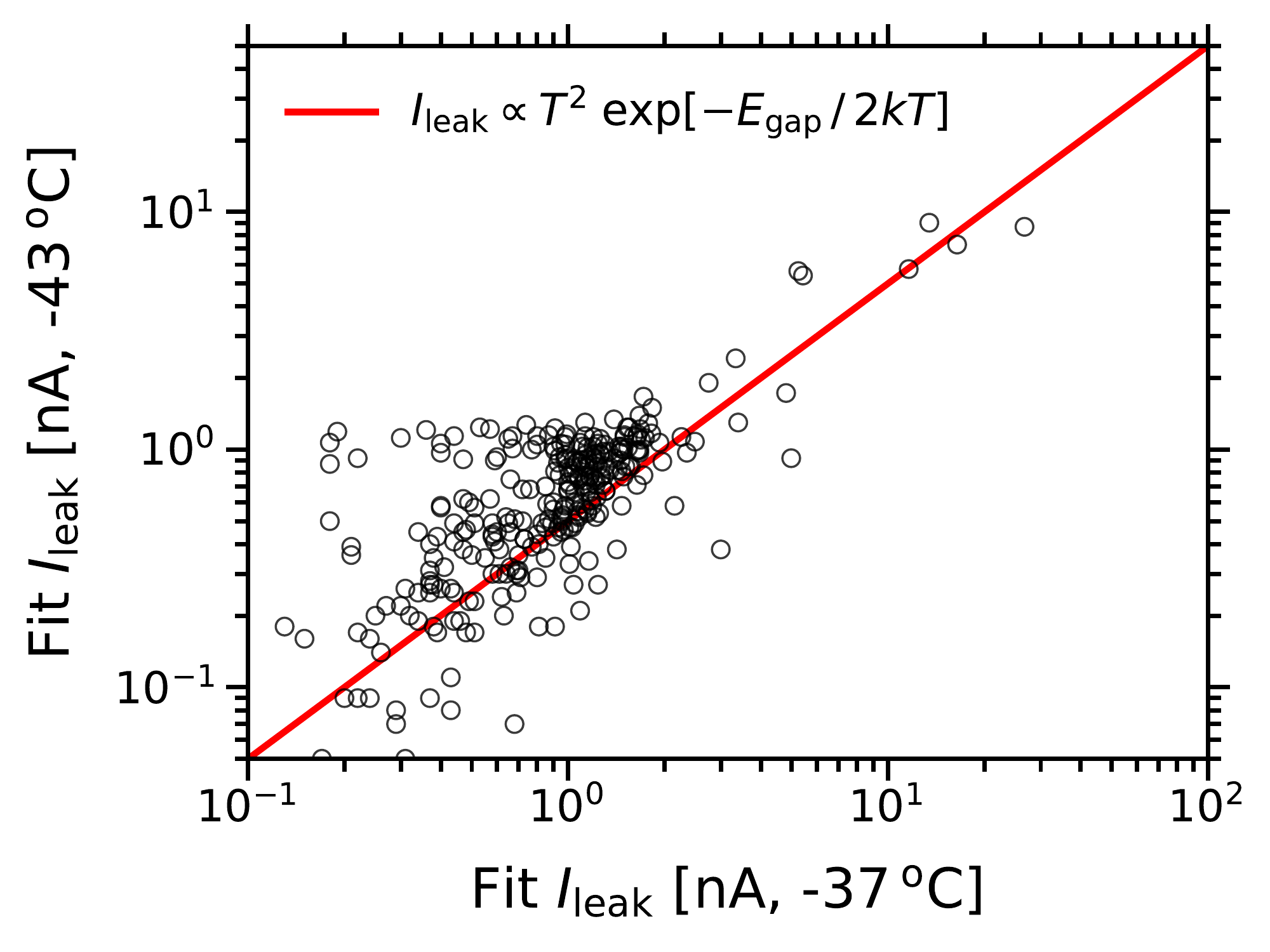}
    }
    
    \subfloat{
        \includegraphics[width=0.95\columnwidth]{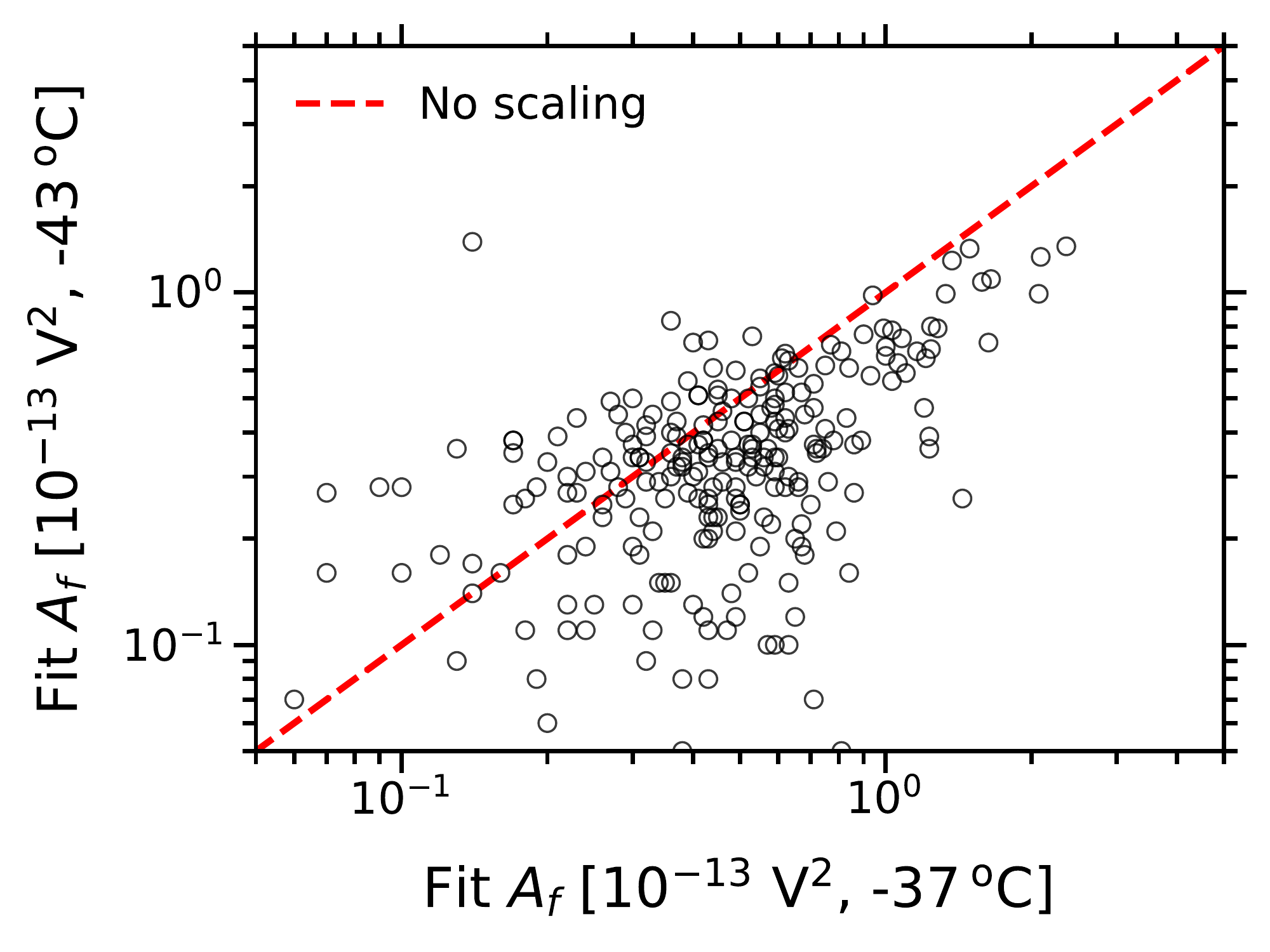}
    }

    \caption{Scatter plots showing the temperature dependence of fit $I_\text{leak}$ (top) and fit $A_f$ (bottom) for Si(Li) strips between --37$\,^\text{o}$C and --43$\,^\text{o}$C. The red line on the $I_\text{leak}$ plot indicates the expected scaling from Eq. (4). The dashed line on the $A_f$ plot indicates $A_f$ constant with temperature.}  
    \label{fig:ileak_af_temperature}
\end{figure}

 Finally, we investigate the scaling of the predicted ASIC x-ray energy resolution with temperature, as we expect there to be range of temperatures at different locations in the payload and at different times during flight. The two detector parameters to which the ASIC resolution is most sensitive are $A_f$ and $I_\text{leak}$. Assuming the leakage current is dominated by the bulk silicon (rather than, e.g., surface effects), we expect $I_\text{leak}$ to scale with absolute temperature $T$ as \cite{Chilingarov_2013}
\begin{equation}
    I_\text{leak}(T) \propto T^2 \exp\left[-\frac{E_\text{gap}}{2kT}\right],
\end{equation}
where $E_\text{gap} \sim 1.14 \text{\,eV}$ is the silicon band-gap energy near $\text{--40}\,^\circ\text{C}$~\cite{Bludau:1974a}. Thus, we expect the leakage current to decrease by a factor ${\sim}$2 when the temperature is reduced by $6\,^\circ\text{C}$ near the GAPS operating temperature of --40$\,^\circ\text{C}$. (We did not have an \textit{a priori} expectation for the temperature dependence of $A_f$.) In Fig.~\ref{fig:ileak_af_temperature} we show the evolution of $I_\text{leak}$ and $A_f$ between $\text{--37}\,^\circ\text{C}$ and --43$\,^\circ\text{C}$ for a sample of 37 detectors. (Due to time constraints, we were unable to test all GAPS detectors at both temperatures.) We note that $I_\text{leak}$ generally scales as expected for values above ${\sim}$5 nA, though this correlation weakens for smaller leakage currents, which we attribute to two sources. First, for low values of $I_\text{leak}$, surface (rather than bulk) currents which do not scale with Eq. (4) may become a significant contribution to the total current. Second, as discussed previously, the noise model of Eq. (1) sometimes has difficulty fitting strips with low values of $I_\text{leak}$, $A_f$, and $R_s$. We observe no significant temperature scaling of $A_f$.
\par To determine the effect of Si(Li) temperature on the predicted ASIC energy resolution, we feed the $I_\text{leak}$ and $A_f$ values extracted at --37$\,^\circ\text{C}$ and --43$\,^\circ\text{C}$ into Eq. (3), and compare with the resolutions expected from $I_\text{leak}$ scaling as Eq. (4), assuming constant $A_f$. We find that the difference in predicted FWHM is ${\lesssim}$0.1 keV. Thus, we conclude that measurement of the operating temperature (faciliated by a sensor integrated into each FEB), combined with the laboratory testing results at $\text{--37}\,^\circ\text{C}$, will allow an adequate prediction of the x-ray energy resolution in flight. Further testing with the full GAPS instrument before flight should help to further confirm this.
\section{\label{sec:conclusion}Conclusion}
The tests described in this work demonstrate that large-area x-ray and charged-particle Si(Li) detectors can be mass-produced and operated above cryogenic temperatures. In particular, we demonstrate that low-noise detector testing systems can be developed in a cost-effective manner using commercial electronics. We find that more than 90\% of the detectors produced for GAPS are operable within the payload, with ${\sim}80\%$ of strips projected to be suitable for both x-ray spectroscopy and charged-particle tracking.
\par With the testing of Si(Li) detectors complete, the assembly of the flight GAPS instrument began in early 2022, and is expected to conclude in mid-2023. Work is ongoing to validate the performance of the full instrument, including the noise contribution of the flight ASIC readout and the combined tracking performance of the Si(Li) tracker and time-of-flight system. The first Antarctic flight of GAPS is currently planned for late 2024.


%



\section*{Acknowledgments}

We thank SUMCO Corporation and Shimadzu Corporation for their cooperation in detector development. We also thank all of the members of the GAPS collaboration for their tireless work and for their comments on this manuscript.

\ifCLASSOPTIONcaptionsoff
  \newpage
\fi



%
\bibliographystyle{IEEEtran}
\bibliography{bib.bib}
%








\end{document}

%% file: sili_table.tex
\begin{table*}[h!]
    \centering
        \caption{\red{Table of criteria and yield rates for x-ray, tracking, and non-usable strips and detectors. The pre-selection criteria are based on RT performance, and all $I_\text{leak}$ values are quoted at --250-V bias. }}

    \begin{tabular}{c|c|c|c|c|c|c}
    \hline
    \hline
        & \multicolumn{3}{c|}{\textbf{Strip-Level}} & \multicolumn{3}{c}{\textbf{Detector-Level}} \\
        \hline
        \textbf{Quality} & \textbf{Pre-selection criteria [RT]} & \textbf{Cold [--37\,$^{\boldsymbol{\circ}}$C]} & \textbf{Yield} & \textbf{Pre-selection criteria [RT]} & \textbf{Cold [--37\,$^{\boldsymbol{\circ}}$C]} & \textbf{Yield} \\
        \hline
        X-ray & $I_\text{leak} < 2\,\upmu\text{A}$ & ASIC FWHM $<$ 4 keV & 7090 & $\ge$7 x-ray strips & $\ge$7 x-ray strips & 776 \\
         & & and $I_\text{leak} < 50\text{\,nA}$ & & and $I_\text{tot} < 300\,\upmu\text{A}$ & and $I_\text{tot} < 1\,\upmu\text{A}$ & \\
        & & & & & & \\
         Tracking & $2\,\upmu\text{A} < I_\text{leak} < 12.5\,\upmu\text{A}$ & ASIC FWHM $>$ 4 keV & 944 & $\le$2 non-usable strips & $\le$3 non-usable strips & 235 \\
         & & and $I_\text{leak} < 50\text{\,nA}$ & & and $I_\text{tot} < 300\,\upmu\text{A}$& and $I_\text{tot} < 1\,\upmu\text{A}$ & \\
         & & & & & & \\
         Non-usable & $I_\text{leak} > 12.5\,\upmu\text{A}$ & Unstable waveform  & 854 & $I_\text{tot} > 300\,\upmu\text{A}$ & $\ge$4 non-usable strips  & 100 \\
         & & or $I_\text{leak} > 50\text{\,nA}$  & & & or $I_\text{tot} > 1\,\upmu\text{A}$ & \\
         \hline
         Total & --- & --- & 8888 & --- & --- & 1111
    \end{tabular}
    \vspace{0.5cm}
    \label{tab:test_results}
\end{table*}